\documentclass[twocolumn,preprintnumbers,amsmath,amssymb,superscriptaddress,prl,showpacs]{revtex4}
\usepackage[english]{babel}
\usepackage[dvips]{graphicx}
\usepackage{dcolumn}
\usepackage{bm}
\usepackage[dvips]{epsfig}

\usepackage{color,ulem}

\begin{document}

\title{Insights on the local dynamics induced by thermal cycling in granular matter}

\author{Baptiste Percier}
\affiliation{Universit\'e de Lyon, Laboratoire de Physique, \'Ecole Normale Sup\'erieure de Lyon, \\
CNRS UMR 5672, 46 All\'ee d'Italie, 69364 Lyon cedex 07, France.}
\author{Thibaut Divoux}
\affiliation{Centre de Recherche Paul Pascal, CNRS UPR 8641, 115 Avenue Schweitzer 33600 Pessac, France.}
\author{Nicolas Taberlet}
\affiliation{Universit\'e de Lyon, Laboratoire de Physique, \'Ecole Normale Sup\'erieure de Lyon, \\
CNRS UMR 5672, 46 All\'ee d'Italie, 69364 Lyon cedex 07, France.}
\affiliation{Universit\'e de Lyon, UFR de physique, Universit\'e Claude Bernard Lyon I}

\date{\today}

\begin{abstract}

In this letter, we report results on the effect of temperature variations on a granular assembly through Molecular Dynamic simulations of a 2D granular column. Periodic dilation of the grains are shown to perfectly mimic such thermal cycling, and allows to rationalize the link between the compaction process, the local grains dynamics and finite size effects. Here we show that the individual grain properties, namely their roughness and elastic modulus define a minimal cycling amplitude of temperature $\Delta T_c$ below which the dynamics is intermittent and spatially heterogeneous while confined into localized regions recently coined ``hot spot" [Amon {\it et al.}, {\it Phys. Rev. Lett.} 108, 135502 (2012)]. Above $\Delta T_c$, the whole column flows while the grains dynamics ranges continuously from cage-like at the bottom of the column to purely diffusive at the top. Our results provide a solid framework for the futur use of thermal cycling as an alternate driving method for soft glassy materials. 
  
\end{abstract}
\maketitle

Flows of soft glassy materials (SGM) such as foams \cite{Kabla:2003,Katgert:2008}, emulsions \cite{Becu:2006}, colloidal glasses \cite{Siebenburger:2012} or granular media \cite{GDR:2004,Forterre:2008} exhibit complex dynamics which common features include aging, spatially inhomogeneous flows (shear-banding) \cite{Ovarlez:2009,Schall:2010}, non-local effects in confined geometries \cite{Mansard:2012}, etc. Recent breakthrough in the field have been twofold. First, dense flows of density matched suspensions and dry granular materials have been described successfully as a dynamical critical phenomenon unifying suspension and granular rheology \cite{Boyer:2011}. Second, the Herschel-Bulkley scaling law, which successfuly accounts for the steady state flow of a large number of SGMs has been linked on the one hand to the material microstructure \cite{Katgert:2009,Bonnecaze:2010}, and on the other hand to the transient yielding process \cite{Divoux:2011}. However, because of subbtle non-local effects and cooperative behaviors \cite{Mansard:2012}, the inertial approach mentionned above fails to describe flows in the limit of vanishingly small shear rates \cite{Forterre:2008}. Furthermore, for such quasistatic flows transient regimes may become extremely long-lived \cite{Divoux:2011}, making irrelevant the use of any constitutive, steady state equations (e.g. the Herschel-Bulkley model) to describe such creeping flows. Consequently, quasitatic flows have triggered a large amount of studies either by continuous shear \cite{Kabla:2003,Divoux:2007} or by periodic shear \cite{Pouliquen:2003,Kabla:2004,Marty:2005,Umbanhowar:2005}, and still urge for more experimental results.

Among the later means, periodic variation of temperature, coined \textit{thermal cycling}, are used to induce slow motion in both thermal \cite{Mazoyer:2006} and athermal systems with granularity \cite{Chen:2006,Divoux:2008}. On top of being a delicate alternative to other driving methods such as cyclic shear \cite{Pouliquen:2003}, tapping \cite{Richard:2005,Umbanhowar:2005} or flow-induced fluidization \cite{Schroter:2005}, this method has successfully brought key information on quasistatic flows of granular matter : grains may exhibit stringlike motions \cite{Slotterback:2008} hinting towards a glassy-like dynamics, while force chains and thus finite size effects are expected to play a key role \cite{Vargas:2001}. However, to date the local grain dynamics has received very little attention and its links to the macroscopic control parameters remains an oustanding issue.

In this letter, we report Molecular Dynamic (MD) simulations of a 2D granular column sumitted to periodic variations of temperature. This thermal cycling is modeled through the quasi-static sinusoidal dilation of grains in a fixed container. This purely geometric approach accounts for the experimental features of previous macroscopic studies and allows to quantify the key role of the grains properties. Here we show that there exists a critical cycling amplitude $\Delta T_c$ fully determined by the grain surface roughness and elastic modulus: for cycling amplitudes lower than $\Delta T_c$, the granular pile compacts by jumps associated with localized plastic events taking place over the whole column. Whereas for cycling amplitudes larger than $\Delta T_c$, the column flows continuously. In this last case, the finite size of the container sets the limit between two different regions: in the upper part of the column, the grains motion is purely diffusive, whereas in the lower part of the column the grains exhibit a glassy-like dynamics including cage rearrangements at short timescales and a diffusive motion at long timescales. Finally, we show that the diffusion coefficient is inversely proportionnal to the pressure inside the column and scales as a non trivial power law with the cycling amplitude and the grain diameter. 

\begin{figure}[t]
\includegraphics[width=0.9\columnwidth]{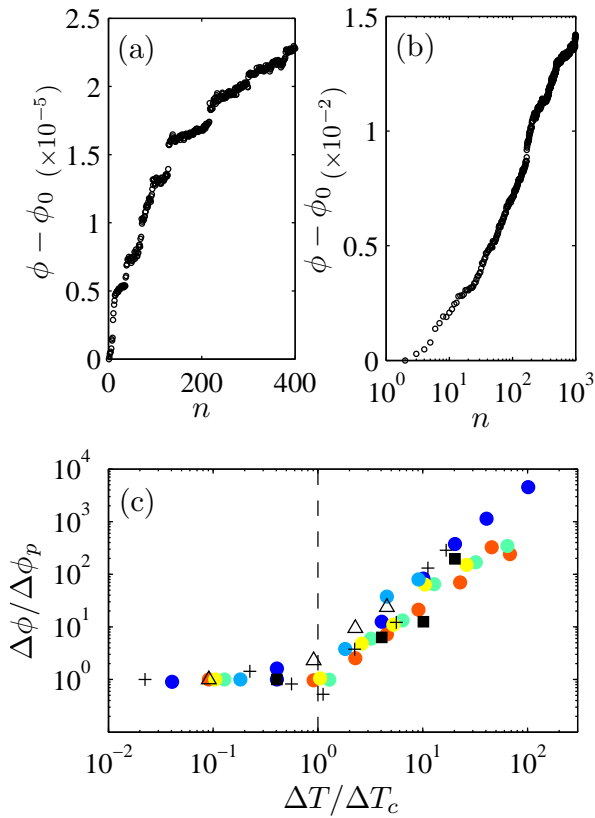}
\caption{Evolution of the packing fraction $\phi-\phi_0$, where $\phi_0=0.81$ is the initial packing fraction, with the number $n$ of cycles of temperature for a cycling amplitude $\Delta T= 10^{-5}$ (a) and $\Delta T= 2.5\times 10^{-3}$ (b). (c) Packing fraction increase $\Delta \phi$ after $n=1000$ cycles normalized by $\Delta \phi_p$ the value of $\Delta \phi$ in the limit of low cycling amplitudes vs. the dimensionless cycling amplitude $\Delta T/\Delta T_c$, where $\Delta T_c$ is the critical cycling amplitude (see text). The symbols correspond to disks of different mechanical properties: [bullet: $k_n=10^5$, $k_t=3300$, $\mu=0.05, 0.25, 0.5, 0.75, 1$]; [$+$: $k_n=10^5$, $k_t=100$; $\mu=0.5$];[$\triangle $: $k_n=10^4$, $k_t=3.3\times 10^4$, $\mu=0.5$]. The vertical dashed line separates the two compaction regimes: by jumps at low $\Delta T$ and continuous for large $\Delta T$.}
\label{figure.1}
\end{figure}
\textit{Set-up.-} MD simulations of $N$ deformable disks ($N=7500$ unless otherwise mentionned) of mean diameter $2R=1$ (20\% polydispersity) and densely packed (initial packing fraction: {$\phi_0 =0.81$) were performed in a rectangular container (width $L=30$, unless otherwise mentionned). Periodic variations of temperature are mimicked through the periodic dilation of the grains by a relative radius $\Delta T$, ranging from $10^{-5}$ to $10^{-2}$. Therefore, the driving is here only geometrical and as such our study strongly differs from previous approaches modelling the temperature field and the heat transfer across the grains \cite{Vargas:2001}. Furthermore, the container is rigid and kept fixed during the cycles of temperature. Thus the grains rearrangements only result from their own dilation and not from the dilation of the boundaries \cite{Clement:1997}. 
\begin{figure}[t]
\includegraphics[width=0.9\columnwidth]{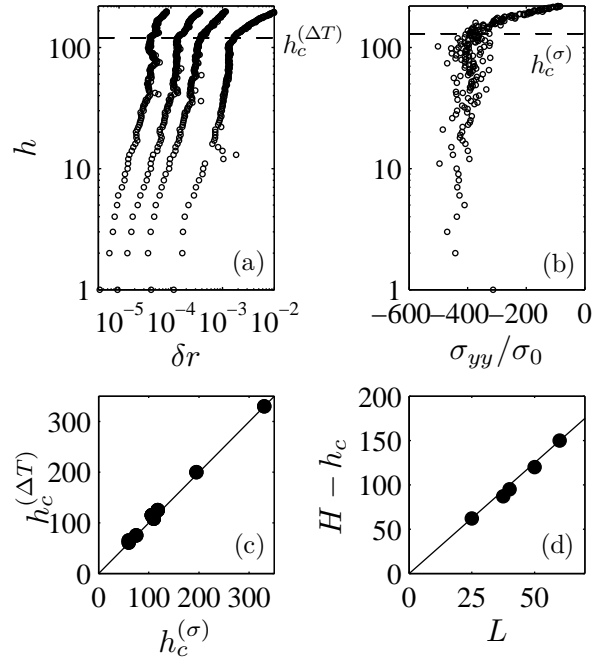}
\caption{(a) Average displacement $\delta r$ of the grains along the column height $h$, for 4 different cyling amplitudes ($\Delta T = 2.5\times 10^{-4},$ $5\times 10^{-4}, 10^{-3}$ and $2.5\times 10^{-3}$ from left to right). (b) Vertical component of the stress tensor $\sigma_{yy}$ averaged over the column width vs. the column height $h$ for $\Delta T=2.5 \times 10^{-3}$. In both (a) and (b), the horizontal dashed line defines a characteristic height $h_c^{(\Delta T)}$ and $h_c^{(\sigma)}$ respectively (see text) which are ploted versus one another in (c) for different parameters (column width: $25 \leq L \leq  60$, and disks number $5.10^3\leq N\leq 2.10^4$). The black line has a slope 1. (d)~Extent of the upper part of the column $H-h_c$ vs. the column width $L$. The black line is the best linear fit: $H-h_c=2.5 L$.}
\label{figure.2}
\end{figure}
Each cycle of temperature is devided into $7.5\times 10^4$ timesteps during which the disks trajectories are obtained by integrating the equations of motion using Verlet's method. Grains are submitted to gravity and contact forces. 
A spring-dashpot scheme is used to model the normal force $F^n_{ij}$ between two contacting disks $i$ and $j$: $F^n_{ij}=2 k_n (r_{\rm eff}/R) \delta + \eta (R/2r_{\rm eff}) \partial_t \delta $, where $\delta $ denotes the disks overlap; $k_n$ the normal stiffness and $r_{eff}^{-1} \equiv r_i^{-1}+r_j^{-1}$ with $r_i$ and $r_j$ the disks radius \cite{Schaeffer:1996}. Unless otherwise mentionned, we choose $k_n=10^5$ and $\eta=100$ leading to a restitution coefficient $e=0.48$, but the results reported in this letter are robust and do not depend on the specific values of these parameters. 
A standard tangential force model with elasticity was implemented \cite{Cundall:1979}: $F^t_{ij}=- \rm sgn ( \zeta_t ) \, min (k_t \zeta_t, \mu F^n_{ij})$ where $\zeta_t = \int_{t_0}^t v_{ij}(t')dt'$ is the net tangential displacement after contact is first established at time $t=t_0$; $v_{ij}$ is the relative velocity; $\mu=0.3$ is the dynamic friction coefficient, and $k_t=3.3\times 10^4$ the spring stiffness. Consequently, the ratio $\mu F_n/k_t$ is much smaller than a disk diameter \cite{Schaeffer:1996}.
Finally, the disks are introduced at the top of the column and left to settle with an additional fluid drag, which is switched off prior to the start of the periodic dilations. Such process produces an initial packing fraction varying from $\phi_0=0.79$ to $\phi_0=0.83$. We carrefully checked that the following results do not depend on $\phi_0$ within this range.

\textit{Global dynamics.-} 
Under periodic dilations the granular column flows in a very similar fashion to what has been previously reported in the literature (fig.~\ref{figure.1}): the packing fraction slowly increases with the number of applied cycles \cite{Chen:2006,Divoux:2008}. Depending on the cycling amplitude $\Delta T$, the compaction process exhibits two different regimes: at low cycling amplitudes, the granular column flows by successive jumps [Fig.~\ref{figure.1}(a)], while the packing fraction remains almost constant. For larger cycling amplitudes, the successive dilations of the disks lead to a slow and continuous compaction of the column which does not reach any steady state, even after $n=1000$ cycles [Fig.~\ref{figure.1}(c)] \cite{remark1}.
In order to quantify these two regimes, the total increase in packing fraction $\Delta \phi= \phi-\phi_0$ after $n=1000$ cycles was computed for various cycling amplitudes $\Delta T$, and disks properties: different elastic modulii ($k_n$) and surface properties ($k_t$, and $\mu$). Surprinsingly, the increase in packing fraction with $\Delta T$ can be scaled onto a single curve while introducing a critical cycling amplitude $\Delta T_c \equiv mg/R \cdot\sqrt{\mu /(k_nk_t)}$ which corresponds to the minimum amplitude to fully mobilize the contact between two disks during a cycle of temperature [Fig.~\ref{figure.1}(b)]. Indeed, at low cycling amplitudes the disks do not slide on each other which reads $k_t R\Delta T<\mu m g$ following Amontons-Coulomb law, and the dilation-induced forces do not overcome gravity: $k_n R\Delta T< mg$. A transition is expected when both disks may slide and the dilation becomes comparable to gravity, which reads: ($k_n R\Delta T$).($k_t R\Delta T$)$=\mu (m g)^2$. As a key result, the amplitude $\Delta T_c$, which quantifies the limit between the two regimes described above, is fully determined by the individual grain properties. 
\begin{figure}[!t]
\includegraphics[width=0.9\columnwidth]{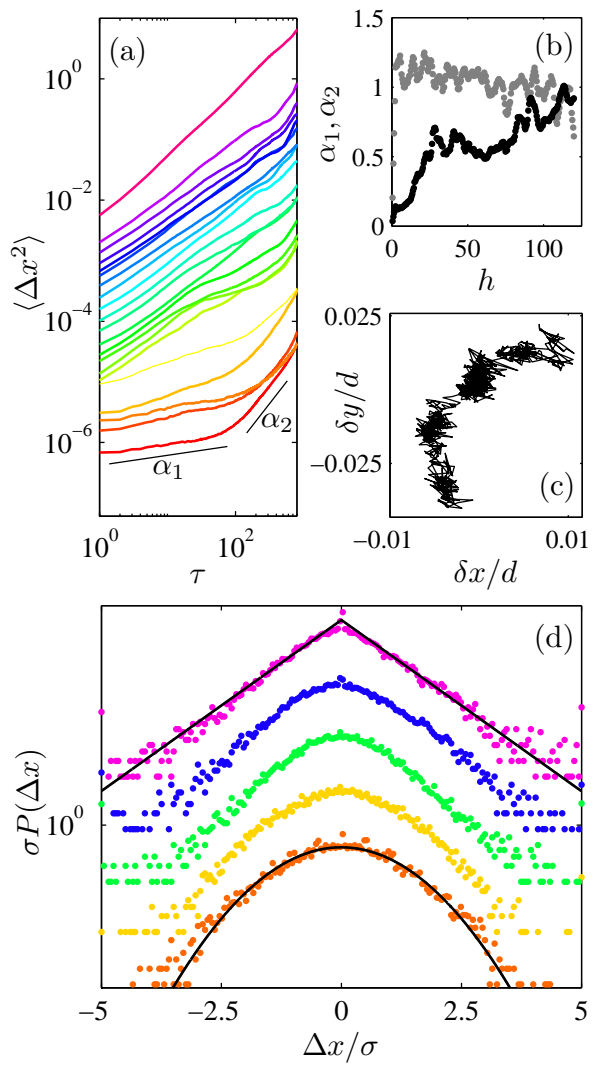}
\caption{ (Color online). (a) Mean square value of the disks horizontal displacement $\langle \Delta x^2 \rangle$ vs. the lagtime $\tau$ in cycles unit at different heights of the column [from $z=$ (bottom, red) to $z=$ (top, pink)]. Each MSD is averaged over an horizontal slice of one disk diameter and fitted by a single or two powerlaws of exponents $\alpha_1$ and $\alpha_2$ at short and long lagtimes respectively. (b)~MSD exponents $\alpha_1$ (black) and $\alpha_2$ (grey) along the column height $h$. (c) Exemple of a disk cagelike trajectory located at the center of the column (d) Normalized probability distribution $\sigma P$ of the horizontal grain displacement $\Delta x$ between two consecutive cycles of temperature, where $\sigma$ stands for the standard deviation of the displacements, computed at different heights of the column: $z=20, 40, 105, 192$ and 240. Each curve is shifted from the previous one by a factor of 10 for the sake of clarity. The lower (upper resp.) fit is a gaussian (exponential resp.) presented as a guide for the eye. $\Delta T=5.10^{-3}>\Delta T_c$. Same color code as in (a).}
\label{figure.3}
\end{figure}

\textit{Confinement effects.-} In order to assess the impact due to the finite size of the system, we compute the average displacement $\delta r$ achieved by the disks in between two successive cycles of temperature along the column height $h$, for different cycling amplitudes below and above $\Delta T_c$ [Fig.~\ref{figure.2}(a)]. Interestingly, the grains displacement is not homogeneous along the vertical, but rather displays two different regimes separated by a characteristic altitude $h_c^{(\Delta T)}$.  
Below $h_c^{(\Delta T)}$, the grains displacement scales as $\sqrt{h}$, and only the prefactor depends on the cycling amplitude $\Delta T$ within the explored range, whereas above $h_c^{(\Delta T)}$, $\delta r$ increases faster with $h$. The existence of these two different regimes is fully accounted for by the friction of the disks on the sidewalls. Indeed, the evolution along the column height of the vertical component of the stress tensor $\sigma_{yy}$ averaged over the column width also exhibits two regimes [Fig.~\ref{figure.2}(b)]. It defines another characteristic height $h_c^{(\sigma)}$ below which the presure is constant, and above which the presure increases strongly with $h$. 
The two characteristic heights $h_c^{(\Delta T)}$ and $h_c^{(\sigma)}$ are in fact equal [Fig.~\ref{figure.2}(c)], which illustrates that the screening effect due to the redirection of the stress, known as the Janssen effect \cite{Vanel:2000}, is responsible for the different grain displacement behavior along the column height. Furthermore, it is worth emphasizing that the size of the upper and unscreened part of the granular column is proportionnal to the column width $L$ [Fig.~\ref{figure.2}(d)], and thus controlled by the finite size of the column. 

\begin{figure}
\includegraphics[width=0.95\columnwidth]{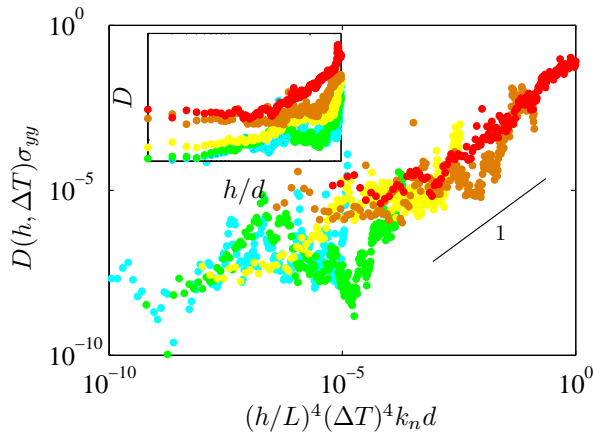}
\caption{(Color online). Diffusion coefficient rescaled by the pressure $D(h,\Delta T)\times \sigma_{yy}$ vs. the column height rescaled by the cycling amplitude $(h\Delta T/L)^4\,k_n	d$. Colors from cyan to red, stand for different cycling amplitudes: from $\Delta T = 2.5\times 10^{-4},$ $5\times 10^{-4}, 10^{-3}$, $2.5\times 10^{-3}$ to $5\times 10^{-3}$. Inset: log-lin plot of the diffusion coefficient $D$ vs. the column height $h/d$.}
\label{figure.4}
\end{figure}

\textit{Local dynamics.-} Let us turn to the individual grain dynamics and start with the case of large cycling amplitudes ($\Delta T>\Delta T_c$). 
To correlate the macroscopic compaction process to the local disks dynamics, we focus on the horizontal displacement (the vertical motion being biased by gravity) of the grains during $\tau$ time steps: $\Delta x(\tau)=x(t+\tau)-x(t)$, and compute its mean square $\langle \Delta x^2 \rangle$ at different altitudes of the column. $\Delta x^2$ is averaged over horizontal slices of one grain diameter and over the duration of the experiment, except for the first twenty cycles which strongly depend on the preparation protocol of the column [Fig.~\ref{figure.3}(a)]. At the bottom of the container, the grains exhibit a subdiffusive behavior at short lag times [$\langle \Delta x(\tau)^2 \rangle \propto \tau^{\alpha_1}$, with $\alpha_1<1$] and a diffusive behavior at larger lag times [$\langle \Delta x(\tau)^2 \rangle \propto \tau^{\alpha_2}$, with $\alpha_2\simeq 1$]. The plateau behavior of the MSD at the bottom of the container is the signature of cage-like rearrangements [Fig.~\ref{figure.3}(c)], as first evidenced in \cite{Weeks:2000} and also reported for other cycling methods in granular matter \cite{Pouliquen:2003,Marty:2005}. From the bottom to the top of the column, one observes a continuous change of the MSD which tends toward a purely diffusive behavior: the exponent $\alpha_1$ increases continuously up to 1 [Fig.~\ref{figure.3}(b)]. 
We can thus define a diffusion coefficient $D \equiv \langle \Delta x^2(\tau) \rangle/ 2\tau$ of the motion induced by thermal cycling, including in the lower part of the column for $\tau \gtrsim 100$ and which a priori depends on the altitude $h$ inside the column, and the cycling amplitude. We found that $D$ increases along the column height, especially above $h_c$ and at a given altitude that $D$ increases with the cyling amplitude [Fig.~\ref{figure.4}~(inset)]. As a central result, both the influence of $\Delta T$ and $h$ on $D$ are captured by the following mastercurve:
\begin{equation}
D(h,\Delta T) \sim  \left(\frac{h}{L}\right)^4 \Delta T^4 \,\frac{k_nd}{\sigma_{yy}}
\end{equation}
The continuous change of the grain dynamics along the column height is also confirmed by computing the probability distribution function of the disks horizontal displacement $P(\Delta x)$, between two successive cycles of temperature [Fig.~\ref{figure.3}(d)]. At the bottom of the column, where the cage-like dynamics takes place, $P(\Delta x)$ is well described by a gaussian: the particles diffuse inside their cage, whereas in the upper part of the column the grains are more mobile and $P(\Delta x)$ follows a decreasing exponential scaling \cite{Chaudhuri:2007}.
\begin{figure}
\includegraphics[width=0.9\columnwidth]{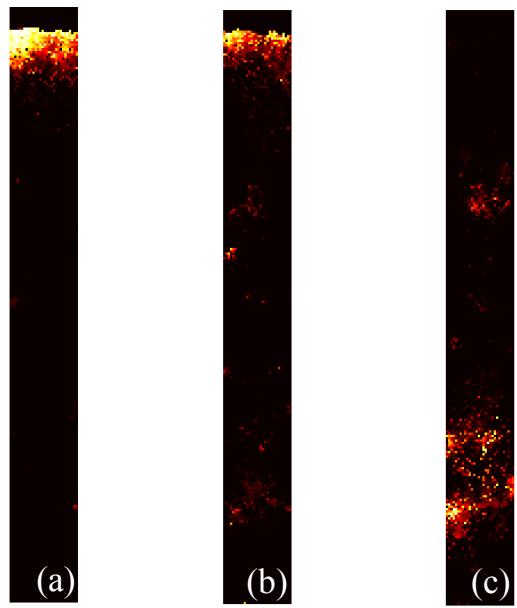}
\caption{(Color online). Colormap for the whole column of the disks displaying the largest displacement, i.e. larger than 5 times the mean value computed every 5 consecutive cycles of temperature, reported for three different cycling amplitudes: $\Delta T=3\Delta T_c$ (a), $\Delta T_c$ (b) and $0.1~\Delta T_c$ (c). For $\Delta T>\Delta T_c$, all the disks move and the ones presenting the largest displacement are located at the top of the column where the pressure is minimum. Whereas, for $\Delta T<\Delta T_c$, the disks presenting the largest displacement are heterogeneously distributed in localized regions inside the column.}
\label{figure.5}
\end{figure}

Finally, focusing on the population of disks which exhibits the largest displacements allows to capture the heterogeneous nature of the dynamics that the diffusion coefficient misses. Indeed, for large cycling amplitudes, all the disks move and those presenting the highest displacement are all located in the upper part of the column  [Fig.~\ref{figure.5}~(a) and the movie in the supplemental material], whereas decreasing the amplitude leads to a very different scenario [Fig.~\ref{figure.5}~(b)]. For $\Delta T< \Delta T_c$, the grains with the largest displacement are heterogeneously spread inside the column into ``hot spots" which temporal dynamics is intermittent [Fig.~\ref{figure.5}~(c) and the movie in the supplemental material]. 

\textit{Discussion.-} 
The master curve reported in figure~\ref{figure.1}(c) demonstrates that the grains surface roughness ($\mu$) and their elastic modulus ($k_t$) play a key and equivalent role on the flow dynamics through a single control parameter $\Delta T_c$. A practical consequence is, for a given amplitude, that smooth and soft grains will be less sensitive to temperature variations than rough and hard grains. In particular, our results rationalize the two regimes of quasistatic flows reported in the literature: isolated rearrangements and continuous compaction \cite{Divoux:2008}. They also explain why lubricated grains experience a faster compaction dynamics than dry grains \cite{Slotterback:2008}. 
The local grain dynamics is rather different on each side of the critical cycling amplitude. Above $\Delta T_c$, finite size effects split the column into two: the upper part where the grains exhibit a diffusive behavior, and the lower part where the grains exhibit cage-like rearrangements at short lagtimes. The diffusion coefficient displays an unusual linear scaling with the disks diameter, which contrasts with the scaling reported for a 2D Couette shear-cell ($d^2$) \cite{Utter:2004}, and that observed in the split bottom cell geometry ($d^3$) \cite{Wandersman:2012}. 
The key difference lays in the scale at which the driving takes place: here, each grain is driven individually by periodic dilation as opposed to any other macroscopic shear where the energy is injected at the macroscopic scale and cascades down towards the grain scale. Nonetheless, the diffusion coefficient is, in a robust fashion, inversly proportionnal to the pressure inside the column as observed in other geometries \cite{Wandersman:2012}. Below $\Delta T_c$, the compaction process is spatially heterogeneous and take place ni localized regions. This work paves the way for a general use of thermal cycling in soft systems to induce controlled and delicate dynamics. Future work will focus on the dynamics of intruders of different mechanical properties.
 

\begin{acknowledgments}
We appreciate useful conversations with B. Blanc, J.-C. G\'eminard, S. Joubaud, M. Leocmach, S. Manneville \& L. Ponson. 
\end{acknowledgments}

\end{document}